\documentclass[peerreview,draftcls,onecolumn,12pt]{IEEEtran}
\usepackage{amsmath,amssymb,latexsym,epsfig,epic}
\usepackage[usenames]{color}
\usepackage{subfigure}
\usepackage{theorem}
\usepackage{relsize}
\usepackage{cite}
\usepackage{booktabs}
\usepackage{graphicx}
\usepackage{epsf}
\usepackage{latexsym}
\usepackage{amsmath}
\usepackage[normalem]{ulem}

\newcommand{\abs}[1]{\left\lvert#1\right\rvert}
\newcommand{\norm}[1]{\left\lVert#1\right\rVert}
\newcommand{\bracket}[1]{\left(#1\right)}

\newcommand{\figref}[1]{Fig.~\ref{#1}}
\newcommand{\tableref}[1]{Table~\ref{#1}}
\newcommand{\secref}[1]{Sec.~\ref{#1}}

\begin{document}

\title{Combining Schemes for Hybrid ARQ with Interference-Aware Successive Decoding
\footnote{The material in this paper was presented in part to 2012 ..., Feb. 2012}}

\author{Hyukjoon~Kwon,~\IEEEmembership{Member,~IEEE},
Jungwon~Lee,~\IEEEmembership{Senior Member,~IEEE},
and Inyup~Kang,~\IEEEmembership{Member,~IEEE}
\thanks{H.~Kwon, J.~Lee and I.~Kang are with Mobile Solution Lab, Samsung US R\&D Center, San Diego, CA, 92130, U.S.A. (e-mail: \{hyukjoon, jungwon\}@alumni.stanford.edu, inyup.kang@samsung.com)}}

\markboth{Submission to IEEE Transactions on Wireless Communications}{Shell \MakeLowercase{\textit{et al.}}: Bare Demo of IEEEtran.cls for Journals}
\maketitle

\begin{abstract}
For decades, cellular networks have greatly evolved to support high data rates over reliable communication. Hybrid automatic-repeat-request (ARQ) is one of the techniques to make such improvement possible. However, this advancement is reduced at the cell edge where interference is not negligible. In order to overcome the challenge at the cell edge, the concept of interference-aware receiver has been recently proposed in which both desired and interference signals are successively decoded, called interference-aware successive decoding (IASD). Although IASD is the advanced receiver technology, interference signals are out of the mobile station's control so that they cannot be requested by the mobile station. For this reason, this paper proposes new combining schemes for the IASD receiver, which operate with hybrid ARQ in a bit level or in a symbol level. In addition, this paper compares the memory requirement among the proposed combining schemes and analyzes the impact of discrete modulation on the proposed scheme. Simulation results presents the superiority of the proposed combining schemes and shows the improvement in terms of the number of transmission.
\end{abstract}

\begin{IEEEkeywords}
MIMO, Interference Mitigation, Interference Decoding, HARQ, IASD
\end{IEEEkeywords}

\clearpage

\section{Introduction} \label{sec:intro}

Modern cellular networks have greatly evolved to achieve high throughput over reliable communication. This advancement of wireless communication enables to support universal frequency reuse and heterogeneous deployments in which pico/femto-cells are covered under a macro cell. Accordingly, cells are easily overlapped so that many cell edge areas are produced. In these areas, interference is not negligible and as a result, the throughput performance is degraded. To address this cell edge problem, a variety of researches have been conducted in the perspective of theory and practice \cite{BJ09,BET11,Jungwon11,Jungwon12}. In theory, \cite{BET11} shows that the capacity region of an interference channel over point-to-point codes can be achieved by either treating interference as noise or jointly decoding both desired and interference signals. In practice, \cite{Jungwon12} has proposed the successive decoding algorithm of both signals for the substitute of joint decoding, called interference-aware successive decoding (IASD). Conventionally, the combining schemes for hybrid automatic-repeat-request (ARQ) do not consider interference signals at a mobile station (MS) because the retransmission request from the MS is only fed back to the desired base station (BS). However, IASD operates to decode interference signals such that the combining schemes for HARQ should be further developed for handling interference signals with hybrid ARQ (HARQ).

The HARQ protocol operates with forward-error-correcting (FEC) codes and ARQ schemes. The FEC codes are used for detecting packet errors and correcting them. Hence, the FEC codes is one of key factors to increase throughput. On the other hand, the ARQ scheme requests to retransmit the data so as to turn the system to be more reliable. Under the HARQ protocol, the received signal at each transmission is not discarded even though it is wrongly detected or decoded. Instead, the erroneous received signal is stored and combined with the signal received at the next transmission.

According to the combining method of the received signals, HARQ schemes can be classified into two types: HARQ with Chase combining (CC) \cite{Chase85} and HARQ with incremental redundancy (IR) \cite{Lin84}. HARQ with CC retransmits the same signal at every transmission so that the diversity gain is expected in time. However, HARQ with IR allows to transmit new information such as parity bits mapped to different positions from the previous transmission. Thus, the bit-level combining can be enabled at both CC and IR schemes since new information at the bit sequence is adjustable for decoding. On the contrary, the symbol-level combining needs to align the symbol vector in order to achieve the combining gain.

So far, many combining schemes for HARQ have been focused on designing the receiver in a point-to-point channel. For example, the signal-to-noise ratio (SNR) is maximized with the matched filter in a single antenna \cite{Haykin} or the maximal-ratio combining (MRC) scheme is used with multiple antennas at the MS \cite{Jakes,Poor}. Under the assumption that channel state information (CSI) is available at the BS, the MRC scheme can be applied at the BS equipped with multiple antennas \cite{Lo99,Kang03,Quan06}. In addition, the MRC scheme is extended to the multi-level weighted combining scheme \cite{Krishnaswamy06}. Although the MRC scheme achieves the throughput gain by maximizing the effective SNR, it could lose a diversity gain by transmitting the signal via the antenna having the best channel condition. The diversity gain for HARQ is analyzed in \cite{Tse00,Feng01,Dighe01}.

As both BS and MS support multiple antennas, the combining schemes for HARQ are also extended to support multiple-input multiple-output (MIMO) systems \cite{Zheng02} and are further improved with a coding system \cite{Koike04}. Also, \cite{Samra06} demonstrates that the bit-to-symbol mapping can be adapted per retransmission in MIMO systems and it outperforms Gray mappings. The combining technique is further developed by incorporating a \emph{turbo principle} by iteratively decoding and exchanging the soft information between the detector and the decoder. This turbo combining technique has been applied from a single antenna system \cite{AitIdir08} to code division multiple access (CDMA) systems with multiple antennas \cite{Chafnaji08}. Moreover, \cite{AitIdir08PIMRC} and \cite{AitIdir10} exploits this technique in the time-domain combining scheme and the frequency-domain combining scheme, respectively. In particular, for the receiver equipped with multiple antennas, it is focused on mitigating inter-stream interference with linear equalizers \cite{Onggosanusi03} or with soft metrics such as log-likelihood ratios (LLR) \cite{Jang09}. However, the transmission from the interference BS with HARQ has not been addressed, yet.

When interference-aware communication is concerned, the HARQ technique needs to be revised since interference is no longer treated as noise. Instead, interference is considered being decodable and its soft information becomes available at the MS such as the IASD receiver. Since interference signals vary at each transmission, it is not expected to achieve the combining gain from interference signals by using conventional HARQ algorithms. Thus, this paper proposes new combining schemes for HARQ handling the decoded interference information at the receiver.

The main contributions of this paper are summarized as follows:
\begin{itemize}
\item proposed the combining schemes for HARQ operating for the IASD receiver.
    \begin{itemize}
    \item the bit-level combining (BLC) directly handles the extrinsic LLRs of the desired signal.
	\item the stacking symbol-level combining (SSLC) is optimal, but requires to increase the memory size and the detector complexity as the number of transmissions increases.
    \item the symbol-level combining with interference cancellation (SLC-IC) is adapted for IASD while maintaining the amount of memory units and the detector complexity.
    \end{itemize}
\item analyzed the amount of memory units required for implementing the proposed schemes
\item analyzed the optimality of the proposed SLC-IC and its decoding performance.
\end{itemize}

This paper is organized as follows: Sec.~\ref{sec:model} describes the system model and Sec.~\ref{sec:IASD} reviews the IASD receiver where the proposed combining schemes operate. In Sec.~\ref{sec:algorithm}, the combining schemes for interference-aware receivers are proposed in a bit level and in a symbol level. Sec.~\ref{sec:complexity} and Sec.~\ref{sec:analysis} analyze the required amount of memory for storing information at each transmission and the decoding performance of the proposed SLC-IC, respectively. Sec.~\ref{sec:result} evaluates the performance and the conclusion is followed in Sec.~\ref{sec:conclusion}.

\section{System Model} \label{sec:model}

This paper considers a Gaussian interference channel where $U$ BSs sends their own messages to the designated MSs. The $u$th BS and its MS are equipped with $N_{t,u}$ transmit antennas and $N_{r,u}$ receive antennas, respectively. A set of messages are divided into $N_{s,u}$ spatial streams and transmitted over $N_{t,u}$ transmit antennas at the BS. When the MS fails to decode messages, it can request to retransmit up to $N$ times. For ease of analysis, $U = 2$ is considered: one is the desired BS and the other is the interference BS. This paper also assumes that the number of transmit antennas and the number of streams are the same, i.e., $N_{t,u} = N_{s,u}$, and these numbers are applied to all BSs. Thus, the subscript $u$ can be omitted in the following.

When HARQ is enabled, the received signal at the $i$th transmission is given by
\begin{eqnarray}
\mathbf{y}_i = \mathbf{H}_{D,i}\mathbf{x}_D + \mathbf{H}_{I,i}\mathbf{x}_{I,i} + \mathbf{n}_i \label{eq:DIeq}
\end{eqnarray}
where the subscript $D$ and $I$ indicate the desired and interference signal, respectively. Hence, $\mathbf{H}_{D,i}, \mathbf{H}_{I,i}, \mathbf{x}_{D}$ and $\mathbf{x}_{I,i}$ denote the channel matrices and the transmitted symbol vectors corresponding to the desired and interference signal, respectively. Since the interference BS is out of the desired MS's control, $\mathbf{x}_{I,i}$ varies at each transmission contrary to $\mathbf{x}_D$. The elements of $\mathbf{x}_{D}$ and $\mathbf{x}_{I,i}$ are modulated with quadrature amplitude modulation (QAM) symbols, each having $N_m$ bits. $\mathbf{n}_i$ is an zero-mean circularly symmetric complex Gaussian (ZMCSCG) noise vector with an identity covariance matrix $\mathbf{I}_{N_r}$. Given the received signal and the channel matrices, the conditional probability distribution function (PDF) of $\mathbf{n}_i$ is expressed as
\begin{eqnarray}
f_{\mathbf{n}_i | \mathbf{H}_{D,i},\mathbf{x}_D,\mathbf{H}_{I,i},\mathbf{x}_{I,i}} \bracket{\mathbf{n}} & \nonumber \\
&\hspace{-3cm}= \frac{1}{\pi^{N_r}}\exp\left( -\norm{\mathbf{y}_i - \mathbf{H}_{D,i}\mathbf{x}_D - \mathbf{H}_{I,i}\mathbf{x}_{I,i}}^2 \right).
\end{eqnarray}
The following section reviews the interference-aware successive decoding (IASD) receiver introduced in \cite{Jungwon12}. The proposed combining schemes for HARQ, which will be described in Sec.~\ref{sec:algorithm}, are assumed to operate at the IASD receiver.

\section{Review of IASD} \label{sec:IASD}

The IASD receiver is designed to effectively mitigate interference signals. Given no knowledge of the interference signal, the simplest method handling interference is to model it as additive Gaussian noise. Then, the interference channel can be converted to a point-to-point channel by whitening interference plus noise \cite{Winters84}. On the other hand, the advanced scheme exploits the fact that interference signals are modulated with $M$-ary QAM symbols which are discretely arranged on the constellation. \cite{Jungwon11} applied the interference-aware detection method by jointly detecting both desired and interference signals with the modulation knowledge of $\mathbf{x}_D$ and $\mathbf{x}_{I,i}$. The most advanced scheme requires the bit level information of $\mathbf{x}_D$ and $\mathbf{x}_{I,i}$. Given the coding information along with the modulation knowledge, both signals can be jointly decoded to resolve the correlation of bit sequences. However, the joint decoding method is extremely complicated because it requires a large computational burden. Hence, \cite{Jungwon12} proposes an alternative scheme that successively decodes both desired and interference signals at the receiver, which is constructed with an iterative detection and decoding (IDD) structure \cite{Hochwald03}.

In this section, the subscript $i$ is temporarily omitted for being consistent with the explanation in \cite{Jungwon12}. As the number of iterations increases, \emph{a priori} information at the detector is updated by using the decoded information as
\begin{align}
L_{m,n}^{(a,1,k)} = \log\frac{P\left(b_{m,n}^k = +1\right)}{P\left(b_{m,n}^k = -1\right)} \label{eq:IASD_apr}
\end{align}
which corresponds to the $m$th bit of the $n$th stream at the $k$th signal. The superscript $k$ can be $D$ or $I$, and is switched to each other by following the detecting order. When $k = D$, \emph{a posteriori} information for the desired signal at the detector is generated as
\begin{align}
L_{m,n}^{(A,1,D)} &= \log\frac{P\left(b_{m,n}^D = +1 | \mathbf{y}\right)}{P\left(b_{m,n}^D = -1 | \mathbf{y}\right)} \\
&\hspace{-0.7cm}= \log\frac
{\sum_{\mathbf{b}^I}\sum_{\substack{\mathbf{b}^D \in \mathbb{B}_{m,n}^{+1}}}P\left(\mathbf{y} | \mathbf{b}^D \mathbf{b}^I \right) P\left(\mathbf{b}^D, \mathbf{b}^I\right)}
{\sum_{\mathbf{b}^I}\sum_{\substack{\mathbf{b}^D \in \mathbb{B}_{m,n}^{-1}}}P\left(\mathbf{y} | \mathbf{b}^D \mathbf{b}^I \right) P\left(\mathbf{b}^D, \mathbf{b}^I\right) } \\
&\hspace{-0.7cm}=\log\sum_{\mathbf{x}^I}\sum_{\substack{\mathbf{x}^D \in \mathbb{X}_{m,n}^{+1}}} \exp\left(\mathfrak{D}_{\mathbf{x}}  + \mathfrak{L}^{(a,1)} \right) \nonumber \\
&\hspace{-0.7cm}- \log\sum_{\mathbf{x}^I}\sum_{\substack{\mathbf{x}^D \in \mathbb{X}_{m,n}^{-1}}}\exp\left(\mathfrak{D}_{\mathbf{x}}  + \mathfrak{L}^{(a,1)} \right) \\
&\hspace{-0.7cm}\stackrel{(a)}{\approx} \max_{\substack{\mathbf{x}^D \in \mathbb{X}_{m,n}^{+1} \mathbf{x}^I}}
\left( \mathfrak{D}_{\mathbf{x}}  + \mathfrak{L}^{(a,1)} \right) \nonumber \\
&\hspace{-0.7cm}- \max_{\substack{\mathbf{x}^D \in \mathbb{X}_{m,n}^{-1}, \mathbf{x}^I}}
\left( \mathfrak{D}_{\mathbf{x}}  + \mathfrak{L}^{(a,1)}\right)  \label{eq:IASD}
\end{align}
where $\mathbf{b}^D$ and $\mathbf{b}^I$ are the bit vector consisting of all bits corresponding to the symbol element of $\mathbf{x}_D$ and $\mathbf{x}_I$, respectively. The Euclidean distance $\mathfrak{D}_{\mathbf{x}}$ and the sum of bit vectors $\mathfrak{L}^{(a,1)}$ using \emph{a priori} information in \eqref{eq:IASD} are given by
\begin{align}
\mathfrak{D}_{\mathbf{x}} &= -\norm{ \mathbf{y} - \mathbf{H}_{D} \mathbf{x}_D - \mathbf{H}_I \mathbf{x}_I}^2 \label{eq:Euclidean} \\
\mathfrak{L}^{(a,1)} &= \frac{1}{2} \mathbf{b}^{I\dagger} \mathbf{L}^{(a,1,I)} + \frac{1}{2} \mathbf{b}^{D\dagger} \mathbf{L}^{(a,1,D)}
\end{align}
where $\mathbf{L}^{(a,1,k)}$ is the vector of LLRs corresponding to $\mathbf{b}^{k}$ with the same superscript. $\mathbb{X}_{m,n}^{b}$ and $\mathbb{B}_{m,n}^{b}$ are the set of transmit symbols and bits where the $m$th bit of the $n$th stream is equal to $b$, which are defined as
\begin{align}
\mathbb{B}_{m,n}^{b} &= \{ \mathbf{b}^D | b_{m,n} = b \} \\
\mathbb{X}_{m,n}^{b} &= \{ \mathbf{x}^D | b_{m,n} = b \}.
\end{align}
The max-log approximation in $(a)$ is used to alleviate the computational burden from the fact that $\log\sum_i\exp x_i \approx \max_i x_i$. Although \eqref{eq:IASD} is developed for the desired signal with $k = D$, it can be expanded for the interference signal with $k = I$ in the same way. Given both \emph{a priori} and \emph{a posteriori} information, the extrinsic information is simply calculated as
\begin{align}
L_{m,n}^{(\textrm{ext},1,k)} = L_{m,n}^{(A,1,k)} - L_{m,n}^{(a,1,k)}.
\end{align}
At each transmission, one of both desired and interference signals is derived depending on the detecting order and is fed into the decoder.

\section{Combining Schemes} \label{sec:algorithm}

This section describes the proposed combining schemes for HARQ in a bit level and in a symbol level. The first scheme is the bit-level combining (BLC) scheme only operating with the LLR of desired signals. The second is the stacking symbol-level combining (SSLC) scheme that is optimal but requires a large amount of memory units and high computational complexity. Lastly, a symbol-level-combining scheme (SLC) with interference-cancellation (IC) is designed to regulate the amount of memory units and its complexity while still achieving the combining gain over BLC.

\subsection{Bit-Level Combining} \label{sec:BLC}

When HARQ is enabled, the same messages for the desired signal are retransmitted on request. Using multiple received signals at the receiver, the BLC scheme applies Bayes' theorem to develop the extrinsic information of the desired signal at the detector. Hence, the extrinsic LLR at the $i$th transmission can be developed into the sum of the individual extrinsic LLR up to the $i$th transmission as follows,
\begin{align}
L_{m,n,i}^{(\textrm{ext},1,D)} &= \log\frac{P\bracket{\{\mathbf{y}_k\}_{k=1}^{i} | b_{m,n}^D = +1}}{P\bracket{\{\mathbf{y}_k\}_{k=1}^{i} | b_{m,n}^D = -1}} \nonumber \\
&= \sum_{k=1}^{i}\log\frac{P\bracket{\mathbf{y}_k | b_{m,n}^D = +1, \{\mathbf{y}_m\}_{m=1}^{k-1}}}{P\bracket{\mathbf{y}_k | b_{m,n}^D = -1, \{\mathbf{y}_m\}_{m=1}^{k-1}}} \nonumber \\
&\stackrel{(b)}{\approx} \sum_{k=1}^{i}\log\frac{P\bracket{\mathbf{y}_k | b_{m,n}^D = +1}}{P\bracket{\mathbf{y}_k | b_{m,n}^D = -1}} \nonumber \\
&=\sum_{k=1}^{i} L_{m,n,k}^{(\textrm{ext},1,D)} \label{eq:BLC}
\end{align}
where $(b)$ is approximated from the fact that the BLC scheme does not store the received signal at the previous transmission. This derivation is only available for the desired signal, not for the interference signal, since $b_{m,n}^D$ is independent of the transmission index $i$.

To implement the BLC scheme, the extrinsic LLRs of the desired signal should be updated at each transmission as in \eqref{eq:BLC}. When the LLR is generated at the detector, it is added to the stored LLR that is the sum of extrinsic LLRs from the $1$st to $(i-1)$th transmission. This value is stored back and is simultaneously conveyed to the decoder after being interleaved at the IASD receiver.

\subsection{Stacking Symbol-Level Combining} \label{sec:ASLC}

Contrary to the BLC scheme, multiple transmission information can be combined at the symbol level. When the amount of memory is extendable and the detector complexity is tolerable, the optimal combining scheme at the $i$th transmission stacks the channel matrices and the received signals up to the $i$th transmission as
\begin{align}
\mathbf{y}_{i}^{\textrm{(s)}} &= \begin{bmatrix} \mathbf{y}_1 \\ \vdots \\ \mathbf{y}_i \end{bmatrix} = \begin{bmatrix} \mathbf{H}_{D,1} & \mathbf{H}_{I,1} & 0 & 0 \\ \vdots & 0 & \ddots & 0 \\ \mathbf{H}_{D,i} & 0 & 0 & \mathbf{H}_{I,i}  \end{bmatrix}\begin{bmatrix} \mathbf{x}_D \\ \mathbf{x}_{I,1} \\ \vdots \\ \mathbf{x}_{I,i} \end{bmatrix} + \begin{bmatrix} \mathbf{n}_1 \\ \vdots \\ \mathbf{n}_i \end{bmatrix} \nonumber \\
&= \mathbf{H}_{D,i}^{\textrm{(s)}}\mathbf{x}_D + \sum_{k=1}^{i}\mathbf{H}_{I,k}^{\textrm{(s)}}\mathbf{x}_{I,k} + \mathbf{n}_{i}^{\textrm{(s)}} \label{eq:optimal}
\end{align}
where $\mathbf{H}_{D,i}^{\textrm{(s)}}$ and $\mathbf{H}_{I,k}^{\textrm{(s)}}$ are the $1$st and $(k+1)$th column submatrices of the concatenated channel matrix, respectively. Then, \eqref{eq:optimal} can be directly applied to the IASD algorithm developed in \eqref{eq:IASD}. Even though the Euclidean distance $\mathfrak{D}_{\mathbf{x}}$ in \eqref{eq:Euclidean} is described with a single interference signal, it can be straightforwardly extended to multiple interference signals by adding more interference signals at $\mathfrak{D}_{\mathbf{x}}$.

To implement the SSLC scheme, the extrinsic LLRs of the desired and multiple interference signals, $L_{m,n,i}^{(\textrm{ext},1,D)}$ and $L_{m,n,k}^{(\textrm{ext},1,I)}$, are sequentially calculated at the $i$th transmission where the transmission index $k$ can be from $1$ to $i$. As following the IASD algorithm, \emph{a priori} informations of $\mathbf{b}^D$ and $\mathbf{b}^{I,k}$ are updated from the decoder. As a result, the SSLC scheme generates the optimal extrinsic LLRs that are obtained from the concatenated signal at the detector.

However, the implementation of the SSLC scheme might be impractical for the following two reasons: the amount of memory units increases in order to concatenate the received signals and all the channel matrices at each transmission. In addition, the detector complexity grows exponentially as the number of transmissions increases.

\subsection{Symbol-Level Combining with IC} \label{sec:SLC_IC}

\begin{figure*}[htb]
\centering
\includegraphics[width=5.5in]{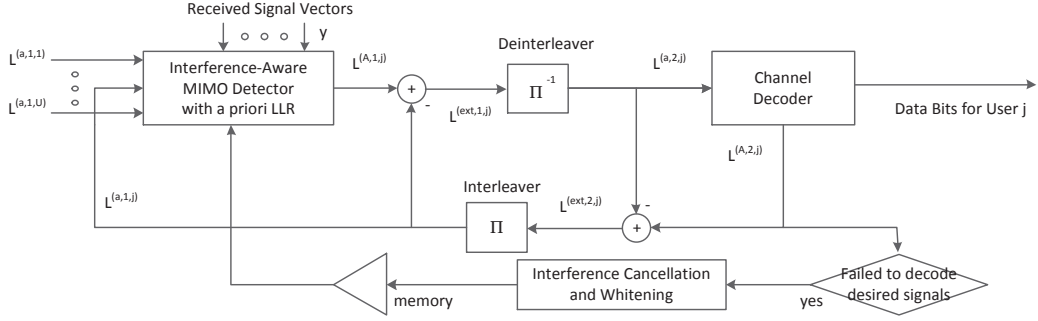}
\caption{The block diagram of the proposed SIC-IC in \secref{sec:SLC_IC} on top of IASD in a $U$-user interference channel}
\label{fig:SLC_IC}
\end{figure*}

This paper proposes the alternative SLC scheme that cancels interference signals of the previous transmission (SLC-IC). For the IC process, the proposed SLC-IC uses the soft information that is already generated at the IASD receiver for its own purpose. The cancellation process enables the SLC-IC scheme to jointly detect only the desired signal and the current interference signal at each transmission. Contrary to the SSLC scheme, the number of symbols used at the detector is maintained so that the computational complexity does not grow even if the number of transmissions increases. \figref{fig:SLC_IC} illustrates a block diagram of the IASD receiver with the implementation of the proposed SLC-IC scheme. Each block is explained in detail below.

At the $i$th transmission, the IASD receiver tries to decode the desired signal in \eqref{eq:DIeq}. When the MS fails to decode the desired signal, the SLC-IC scheme cancels the interference signal before receiving the $(i+1)$th transmission. Then, the received signal where interference is cancelled is given by
\begin{align}
\acute{\mathbf{y}}_i &= \mathbf{y}_i - \mathbf{H}_{I,i}\bar{\mathbf{x}}_{I,i} \\
&= \mathbf{H}_{D,i}\mathbf{x}_D + \mathbf{H}_{I,i}\bracket{\mathbf{x}_{I,i} - \bar{\mathbf{x}}_{I,i}} + \mathbf{n}_i \label{eq:IC} \\
&= \mathbf{H}_{D,i}\mathbf{x}_D + \mathbf{v}_i \label{eq:residual}
\end{align}
where $\mathbf{v}_i$ consists of the residual interference and the received noise. $\bar{\mathbf{x}}_{I,i}$ is the soft estimate of the interference signal vector and its $n$th scalar element is expressed as
\begin{align}
\bar{x}_{I,i}(n) &= E\left[ x_{I,i}(n) | \mathbf{L}^{(A,2,I)} \right] \nonumber \\
&\hspace{-.5cm}= \sum_{s_k \in \mathbb{C}}s_k \prod_{m=1}^{M}\frac{1}{2}\bracket{1 + b_{m,n}^{I}\tanh\bracket{\frac{L_{m,n,i}^{(A,2,I)}}{2}}} \label{eq:bar_x_I}
\end{align}
where $\mathbb{C}$ refers to the set of $M$-ary QAM constellation points. $L_{m,n,i}^{(A,2,I)}$ denotes \emph{a posteriori} LLR of the interference signal corresponding to the $m$th bit of the $n$th symbol at the $i$th transmission, obtained at the decoder. Although the interference signal $\mathbf{x}_{I,i}$ is canceled by $\bar{\mathbf{x}}_{I,i}$, it is not guaranteed that the interference signal is perfectly canceled. As a result, the residual interference signal still remains in \eqref{eq:IC}, which prevents the desired signal from being decoded correctly at the next transmission. Therefore, the SLC-IC scheme whitens the residual interference plus noise, and transforms \eqref{eq:IC} into a point-to-point channel for the desired signal as
\begin{align}
\tilde{\mathbf{y}}_i &= \mathbf{R}_{I,i}^{-\frac{1}{2}}\acute{\mathbf{y}}_i \label{eq:whitening} \\
&= \tilde{\mathbf{H}}_{D,i}\mathbf{x}_D + \tilde{\mathbf{n}}_{i} \label{eq:whitened}
\end{align}
where the covariance matrix of the residual interference with the received noise, $\mathbf{v}_i$, in \eqref{eq:residual} is given by
\begin{align}
\mathbf{R}_{I,i} = \mathbf{H}_{I,i}\mathbf{Q}_{I,i}\mathbf{H}_{I,i}^{\dagger} + \mathbf{I}_{N_r}. \label{eq:SLR_R}
\end{align}
The diagonal matrix $\mathbf{Q}_{I,i}$ is the soft covariance matrix for the residual interference, and is correspondingly defined as $E\left[\bracket{\mathbf{x}_{I,i} - \bar{\mathbf{x}}_{I,i}}\bracket{\mathbf{x}_{I,i} - \bar{\mathbf{x}}_{I,i}}^*\right]$. The $n$th diagonal element of $\mathbf{Q}_{I,i}$ is expressed as
\begin{align}
\mathbf{Q}_{I,i}(n,n) = E\left[ \abs{x_{I,i}(n)}^2 | \mathbf{L}^{(A,2,I)} \right] - \abs{\bar{x}_{I,i}(n)}^2.
\end{align}
Similar to \eqref{eq:bar_x_I}, the $2$nd order expectation of $\abs{x_{I,i}(n)}$ can be calculated in the same way by using \emph{a posteriori} information at the decoder as
\begin{align}
E\left[ \abs{x_{I,i}(n)}^2 | \mathbf{L}^{(A,2,I)} \right] &\nonumber \\
&\hspace{-3cm}= \sum_{s_k \in \mathbb{C}}\abs{s_k}^2 \prod_{m=1}^{M}\frac{1}{2}\bracket{1 + b_{m,n}^{I}\tanh\bracket{\frac{L_{m,n,i}^{(A,2,I)}}{2}}}. \label{eq:bar2_x_I}
\end{align}
Then, the received signal $\tilde{\mathbf{y}}_i$ and the desired channel $\tilde{\mathbf{H}}_{D,i}$ in the transformed channel of \eqref{eq:whitened} are added up with the corresponding signals at the previous transmission, which is loaded at the memory in \figref{fig:SLC_IC}. As multiplying the Hermitian of the desired channel $\tilde{\mathbf{H}}_{D,i}$, both signals are updated with the maximal-ratio combining (MRC) technique \cite{Jang09}. Consequently, the updated signals up to the $i$th transmission can be expressed with the sum of the current signal and the previously stored information as
\begin{align}
\hat{\mathbf{y}}_i  &= \tilde{\mathbf{H}}_{D,i}^\dagger\tilde{\mathbf{y}}_i + \sum_{k=1}^{i-1}\tilde{\mathbf{H}}_{D,k}^\dagger\tilde{\mathbf{y}}_k \label{eq:SLC_y}\\
\hat{\mathbf{H}}_i  &= \tilde{\mathbf{H}}_{D,i}^\dagger\tilde{\mathbf{H}}_{D,i} + \sum_{k=1}^{i-1}\tilde{\mathbf{H}}_{D,k}^\dagger\tilde{\mathbf{H}}_{D,k} \label{eq:SLC_H}
\end{align}
which are stored back to the memory for assisting to decode at the next transmission. In the updating process, the noise signals are also added up to $\hat{\mathbf{n}}_i = \sum_{k=1}^{i}\tilde{\mathbf{H}}_{D,k}\tilde{\mathbf{n}}_k$. Since each transmission is independent, the covariance matrix of $\hat{\mathbf{n}}_i$ is the same as $\hat{\mathbf{R}}_i = \hat{\mathbf{H}}_i$ so that no additional memory is required for storing $\hat{\mathbf{R}}_i$.

At the $(i+1)$th transmission, the received signal $\mathbf{y}_{i+1}$ is combined with $\hat{\mathbf{y}}_{k}$ previously stored up to the $i$th transmission as follows,
\begin{align}
\breve{\mathbf{y}}_{i+1} &= \begin{bmatrix} \hat{\mathbf{H}}_{i}^{-\frac{1}{2}}\hat{\mathbf{y}}_i \\ \mathbf{y}_{i+1} \end{bmatrix} \\
&= \begin{bmatrix} \hat{\mathbf{H}}_{i}^{\frac{1}{2}} & 0 \\ \mathbf{H}_{D,i+1} & \mathbf{H}_{I,i+1} \end{bmatrix} \begin{bmatrix} \mathbf{x}_D \\ \mathbf{x}_{I,{i+1}} \end{bmatrix} + \begin{bmatrix} \hat{\mathbf{n}}_{i} \\ \mathbf{n}_{i+1} \end{bmatrix} \\
&= \breve{\mathbf{H}}_{D,i+1}\mathbf{x}_D + \breve{\mathbf{H}}_{I,i+1}\mathbf{x}_{I,i+1} + \breve{\mathbf{n}}_{i+1}.
\end{align}
Since $\hat{\mathbf{y}}_i$ is a colored signal with the covariance $\hat{\mathbf{H}}_i$, it is pre-whitened and combined with $\mathbf{y}_{i+1}$. Moreover, the combined signal $\breve{\mathbf{y}}_{i+1}$ has the same format as \eqref{eq:DIeq} so that no additional efforts to change the IASD structure are required except for installing the memory block and the interference cancellation block.

To implement the SLC-IC scheme, after the $i$th transmission, two parameters in \eqref{eq:SLC_y} and \eqref{eq:SLC_H} need to be updated by using the decoded bit sequence information at the IASD receiver. When the $(i+1)$th transmission is requested, the MS loads the store signals and transforms the channel model to \eqref{eq:whitened}. With combination of current signals, the IASD receiver achieves the HARQ gain at the symbol level.

\section{Memory Requirement} \label{sec:complexity}

This section evaluates the amount of memory required for implementing the proposed BLC, SSLC and SLC-IC schemes. A single unit of memory denotes the memory size needed to store one real number. \tableref{tbl:memory} tabulates the amount of memory units that each scheme requires for the detection at the $i$th transmission. Except for the SSLC scheme, the memory size of both the BLC and SLC-IC schemes is not a function of the number of retransmissions.

The BLC scheme only needs to store the extrinsic LLR of the desired signal. For $M$-ary QAM, there are $2^{N_m}$ constellation points per symbol so that $N_m$ units of memory is needed to store one QAM symbol. Thus, the total number of memory units for $N_s$ spatial streams is simply $N_m N_s$.

In case of the SSLC scheme, the concatenated channel matrix grows with respect to the number of retransmissions as in \eqref{eq:optimal}. Thus, the required amount of memory units varies as the number of retransmissions increases. The channel matrix is independent at each transmission and its size is $N_r \times N_s$. Since the elements of $\mathbf{H}_{D,i}$ or $\mathbf{H}_{I,i}$ are complex numbers, two units per number are required. Hence, prior to requesting the $i$th transmission, the optimal scheme requires a total of $2(i-1)N_sN_r$ memory units to store the concatenated channel matrix.

Regarding the SLC-IC scheme, the complex received signal $\hat{\mathbf{y}}_i$ in \eqref{eq:SLC_y} requires $2N_s$ memory units. Since the size of the desired channel matrix in \eqref{eq:SLC_H} is $N_s \times N_s$, it would need $2N_s^2$ memory units. However, using the fact that $\hat{\mathbf{H}}_i$ is conjugate symmetric, the amount of memory units for storing the desired channel can be reduced more. First, the diagonal elements of $\hat{\mathbf{H}}_i$ are real numbers so that only $N_s$ memory units are required. Then, the rest are conjugate symmetric with respect to the diagonal. Therefore, $(N_s^2 - N_s)/2$ elements are enough to be stored. Since each element is complex, the off-diagonal elements require $N_s^2 - N_s$ memory units. In total, the memory requirement for the SLC-IC scheme is $N_s^2 + 2N_s$.

\begin{table}[htb]
\caption{ Memory Requirement of Combining Schemes } \label{tbl:memory}
\vspace{-.2cm}
\begin{center}
\begin{tabular}{ ll }
\toprule
Combining Scheme & Memory Unit \\ \midrule \midrule
BLC   & $ N_m N_s $ \\ \midrule
SSLC   & $ 2(i-1)N_sN_r $ \\ \midrule
SLC-IC  & $ (N_s + 2)N_s $  \\
\bottomrule
\end{tabular}
\end{center}
\end{table}

In practice, the required amount of memory units highly depends on the hardware implementation. Accordingly, the cost per memory unit also varies. Suppose that $N_s = N_t = 2$ and $4$ QAM is used. Then, the BLC and SLC-IC schemes need $4$ and $8$ memory units, respectively. When $16$ QAM is used, both schemes equally need $8$ memory units. Although the SLC-IC scheme could spend more units compared to the BLC scheme, it achieves the performance advantage over the BLC scheme as shown in \secref{sec:result}.


\section{Analysis on the SLC-IC scheme} \label{sec:analysis}

This section examines the optimality of the SLC-IC scheme under the assumption that interference is perfectly canceled. This assumption can be generalized that the residual interference follows a Gaussian distribution. Then, since the received noise is also a complex Gaussian random vector, the residual interference plus noise has a Gaussian distribution. As a result, the whitening operation in \eqref{eq:whitening} loses no information. On the other hand, this section analyzes the SLC-IC scheme where interference is modulated on the discrete constellation. In this case, the residual interference plus noise is not Gaussian but just treated as Gaussian. Hence, it becomes a factor to degrade the performance.

\subsection{Gaussian Approximation}

Suppose that the residual interference, $\mathbf{x}_{I,k} - \bar{\mathbf{x}}_{I,k} \: \forall k = 1, \ldots, i-1$, is Gaussian distributed. Then, the residual interference plus noise, $\mathbf{v}_k$ in \eqref{eq:residual}, also follows a Gaussian distribution with mean zero and the covariance $\mathbf{R}_{I,k}$.  Let $\mathbf{x} \sim \mathcal{N}\bracket{\mu, \Sigma}$ for $\mu \in \mathbf{S}^{N_r \times 1}$ and $\Sigma \in \mathbf{S}^{N_r \times N_r}_{++}$. Since a Gaussian random variable holds the multiplication property
\begin{align}
\mathbf{A}\mathbf{x} \sim \mathcal{N}\bracket{\mathbf{A}\mu, \mathbf{A} \Sigma \mathbf{A}^\dagger}
\end{align}
for a matrix $\mathbf{A} \in \mathbf{S}^{N_r \times N_r}$, the vector $\mathbf{v}_k$ can be transformed to have the standard normal distribution by multiplying $\mathbf{R}_{I,k}^{-\frac{1}{2}}$. Under the procedure, the interference whitening operation in \eqref{eq:whitening} loses no information for obtaining $\tilde{\mathbf{H}}_{D_k}$. In this scenario, the optimal combining scheme is achieved by assembling all the previous channel information, $\tilde{\mathbf{H}}_{D_k}$ where $k \leq i-1$, and incorporating them with the $i$th transmission information. Then, the concatenated signal forms a channel model as
\begin{align}
\mathbf{y}_{i}^{\textrm{(g)}} = \begin{bmatrix} \tilde{\mathbf{y}}_1 \\ \vdots \\ \tilde{\mathbf{y}}_{i-1} \\ \mathbf{y}_{i} \end{bmatrix}
= \begin{bmatrix} \tilde{\mathbf{H}}_{D,1} & 0 \\ \vdots & \vdots \\ \tilde{\mathbf{H}}_{D,i-1} & 0 \\ \mathbf{H}_{D,i} & \mathbf{H}_{I,i} \end{bmatrix}\begin{bmatrix} \mathbf{x}_D \\ \mathbf{x}_{I,i} \end{bmatrix}
+ \begin{bmatrix} \tilde{\mathbf{n}}_1 \\ \vdots \\ \tilde{\mathbf{n}}_{i-1} \\ \mathbf{n}_{i} \end{bmatrix}
\end{align}
which can be equivalently rewritten to
\begin{align}
\mathbf{y}_{i}^{\textrm{(g)}} &= \begin{bmatrix} \tilde{\mathbf{y}}_{1:i-1}^{\textrm{(p)}} \\ \mathbf{y}_{i+1} \end{bmatrix} = \begin{bmatrix} \tilde{\mathbf{H}}_{1:i-1}^{\textrm{(p)}} & 0 \\ \mathbf{H}_{D,i} & \mathbf{H}_{I,i} \end{bmatrix} \begin{bmatrix} \mathbf{x}_D \\ \mathbf{x}_{I,{i}} \end{bmatrix} + \begin{bmatrix} \tilde{\mathbf{n}}_{1:i-1}^{\textrm{(p)}} \\ \mathbf{n}_{i} \end{bmatrix} \label{eq:SLC_IC_Gaussian1} \\
&= \mathbf{H}_{D,i}^{\textrm{(g)}}\mathbf{x}_{D} + \mathbf{H}_{I,i}^{\textrm{(g)}}\mathbf{x}_{I,i} + \mathbf{n}_{i}^{\textrm{(g)}} \label{eq:SLC_IC_Gaussian2}
\end{align}
where $\tilde{\mathbf{y}}_{1:i-1}^{\textrm{(p)}}$, $\tilde{\mathbf{H}}_{1:i-1}^{\textrm{(p)}}$ and $\tilde{\mathbf{n}}_{1:i-1}^{\textrm{(p)}}$ are the combined received signal, the combined desired channel and the combined noise vector up to $(i-1)$th transmission, respectively. Consequently, $\mathbf{H}_{D,i}^{\textrm{(g)}}$ and $\mathbf{H}_{I,i}^{\textrm{(g)}}$ denote the first and second column matrices of the channel matrix in \eqref{eq:SLC_IC_Gaussian1}.

In \cite[Theorem 1]{Jang09}, it is proved that the MRC scheme for the SLC of HARQ is equivalent to the optimal scheme in a point-to-point MIMO channel. This theorem can be applied to \eqref{eq:SLC_IC_Gaussian1} because no interference signals up to the $(i-1)$th transmission are decoded, again. When the $i$th interference signal is included, the joint detection with $\tilde{\mathbf{H}}_{1:i-1}^{\textrm{(p)}}$ still maintains the optimality.

The optimal LLR value under the Gaussian assumption can be derived as follows. At each transmission, \emph{a priori} information is not stored but initialized as zero. Thus, the combining occurs without \emph{a priori} information at the first iteration. From \eqref{eq:SLC_IC_Gaussian2}, the LLR is developed into
\begin{align}
L_{m,n}^{(A,1,D)} 
&= \log\frac
{\sum_{\mathbf{b}^I}\sum_{\substack{\mathbf{b}^D \in \mathbb{B}_{m,n}^{+1}}}P\left(\mathbf{y}_{i}^{\textrm{(g)}} | \mathbf{b}^D \mathbf{b}^I \right)}
{\sum_{\mathbf{b}^I}\sum_{\substack{\mathbf{b}^D \in \mathbb{B}_{m,n}^{-1}}}P\left(\mathbf{y}_{i}^{\textrm{(g)}} | \mathbf{b}^D \mathbf{b}^I \right)}  \label{eq:LLRgau1}\\
&\hspace{-1cm}=\log\frac{\sum_{\mathbf{x}_{I,i}}\sum_{\substack{\mathbf{x}_D \in \mathbb{X}_{m,n}^{+1}}} \exp\left(\mathfrak{D}_{\mathbf{x},i}^{\textrm{(g)}} \right)}{\sum_{\mathbf{x}_{I,i}}\sum_{\substack{\mathbf{x}_D \in \mathbb{X}_{m,n}^{-1}}}\exp\left(\mathfrak{D}_{\mathbf{x},i}^{\textrm{(g)}} \right)} \label{eq:LLRgau2} \\
&\hspace{-1cm}\approx \max_{\substack{\mathbf{x}_D \in \mathbb{X}_{m,n}^{+1}, \mathbf{x}_{I,i}}}
\mathfrak{D}_{\mathbf{x},i}^{\textrm{(g)}} - \max_{\substack{\mathbf{x}_D \in \mathbb{X}_{m,n}^{-1}, \mathbf{x}_{I,i}}}
\mathfrak{D}_{\mathbf{x},i}^{\textrm{(g)}} \label{eq:LLRgau3}\\
&\hspace{-1cm}= \bracket{\max_{\substack{\mathbf{x}_D \in \mathbb{X}_{m,n}^{+1}, \mathbf{x}_{I,i}}}
\mathfrak{D}_{\mathbf{x},i} - \max_{\substack{\mathbf{x}_D \in \mathbb{X}_{m,n}^{-1}, \mathbf{x}_{I,i}}}
\mathfrak{D}_{\mathbf{x},i}}  \nonumber \\
&\hspace{-1cm}+ \bracket{\max_{\substack{\mathbf{x}_D \in \mathbb{X}_{m,n}^{+1}}}
\sum_{k=1}^{i-1}\tilde{\mathfrak{D}}_{\mathbf{x},k} - \max_{\substack{\mathbf{x}_D \in \mathbb{X}_{m,n}^{-1}}}
\sum_{k=1}^{i-1}\tilde{\mathfrak{D}}_{\mathbf{x},k}} \label{eq:LLRgau4}
\end{align}
where the Euclidean distance variables used in \eqref{eq:LLRgau2}, \eqref{eq:LLRgau3} and \eqref{eq:LLRgau4} are defined as
\begin{align}
\mathfrak{D}_{\mathbf{x},i}^{\textrm{(g)}} &= -\norm{ \mathbf{y}_{i}^{\textrm{(g)}} - \mathbf{H}_{D,i}^{\textrm{(g)}} \mathbf{x}_D - \mathbf{H}_{I,i}^{\textrm{(g)}} \mathbf{x}_{I,i}}^2 \\
&= \mathfrak{D}_{\mathbf{x},i} + \sum_{k=1}^{i-1}\tilde{\mathfrak{D}}_{\mathbf{x},k}, \\
\tilde{\mathfrak{D}}_{\mathbf{x},k} &= -\norm{\tilde{\mathbf{y}}_{k} - \tilde{\mathbf{H}}_{D,k}\mathbf{x}_D}^2, \label{eq:tilde_Euclidean} \\
\mathfrak{D}_{\mathbf{x},i} &= -\norm{\mathbf{y}_i - \mathbf{H}_{D,i}\mathbf{x}_D - \mathbf{H}_{I,i} \mathbf{x}_{I,i}}^2.
\end{align}
When \emph{a priori} information is updated at the IASD receiver, this development can be directly extended by adding $\mathfrak{L}^{(a,1)}$ to $\mathfrak{D}_{\mathbf{x},i}^{\textrm{(g)}}$ in \eqref{eq:LLRgau2}.

\subsection{Discrete Modulation}

In practice, it is less likely that interference signals are perfectly decoded due to lack of the number of iterations. Thus, the residual interference does not follow a Gaussian distribution but follows a Gaussian mixture distribution as explained in Appendix~\ref{sec:appendix1}. Accordingly, the PDF of $\mathbf{v}_{i}$ should be obtained from a Gaussian mixture distribution. Thus, the denominator and numerator in the LLR equation need to be modified for the accurate soft information. Given \emph{a posteriori} information of the interference signal, the denominator of the accurate LLR for \eqref{eq:residual} should have a form of
\begin{align}
\sum_{\substack{\mathbf{x}_D \in \mathbb{X}_{m,n}^{-1}}}\sum_{\mathbf{s}_k \in \mathbb{C}} \exp\bracket{\mathfrak{D}_{\mathbf{x},i,k}^{\textrm{mix}} + \log w_{i,k}} \label{eq:Gaussian_Mixture}
\end{align}
where
\begin{align}
\mathfrak{D}_{\mathbf{x},i,k}^{\textrm{mix}} &= -\norm{\mathbf{y}_i - \mathbf{H}_{D,i}\mathbf{x}_D - \mathbf{H}_{I,i}\bracket{\mathbf{s}_k - \bar{\mathbf{x}}_{I,i}}}^2. \\
w_{i,k} &= P_{\mathbf{x}_I}\bracket{\mathbf{x}_{I,i} = \mathbf{s}_k}
\end{align}
As defined in Appendix~\ref{sec:appendix1}, $w_{i,k}$ is the symbol probability that the interference signal vector $\mathbf{x}_{I,i}$ is the same as the symbol vector $\mathbf{s}_k$ whose element is the point on the $M$-ary QAM constellation. $k$ can be $1$ to $M^{N_s}$. However, by treating $\mathbf{v}_i$ as Gaussian, the interference whitening operation is performed and the conventional LLR is used with the denominator as
\begin{align}
\sum_{\substack{\mathbf{x}_D \in \mathbb{X}_{m,n}^{-1}}} \exp\bracket{\tilde{\mathfrak{D}}_{\mathbf{x},i}} \label{eq:Gaussian_approximation}
\end{align}
where the Euclidean distance $\tilde{\mathfrak{D}}_{\mathbf{x},i}$ is defined in \eqref{eq:tilde_Euclidean}. The discrepancy between these distributions could affect the performance of the SLC-IC scheme.

\begin{figure}[htb]
\centering
\includegraphics[width=3.2in]{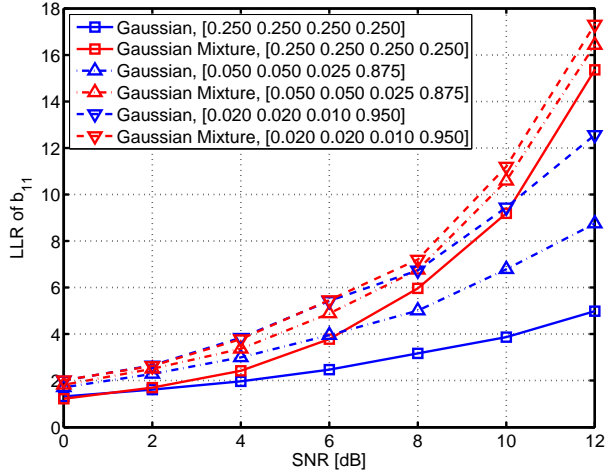}
\caption{The LLRs in \eqref{eq:Gaussian_Mixture} and \eqref{eq:Gaussian_approximation} are compared at $4$ QAM and SIR $0$ dB. It is assumed that \emph{a posteriori} LLRs provide the symbol probability of interference signals with $[0.25 \: 0.25 \: 0.25 \: 0.25]$, $[0.05 \: 0.05 \: 0.025 \: 0.875]$, and $[0.02 \: 0.02 \: 0.01 \: 0.95]$, respectively. $N_r = N_t = N_s = 2$.}
\label{fig:analysis_discrete}
\end{figure}

\figref{fig:analysis_discrete} shows the LLRs of $b_{1,1}^{D}$ that are calculated by \eqref{eq:Gaussian_Mixture} and \eqref{eq:Gaussian_approximation} with different \emph{a posteriori} information for $b_{m,n}^{I}$ at the SIR of $0$ dB. The absolute value of LLRs denotes how much they are reliable regardless of the bit sign. In this sense, \figref{fig:analysis_discrete} indicates that the interference whitening operation by treating $\mathbf{v}_i$ as Gaussian could degrade the performance. If interference is poorly decoded, the performance gap would be wider such as the plots with equal symbol probabilities. If interference is well decoded, the symbol probabilities become severely asymmetric and $\mathbf{x}_{I,k} - \bar{\mathbf{x}}_{I,k}$ is close to zero. Then, the performance gap is accordingly reduced. In particular, the SLC-IC scheme is likely to operate at low SNR. Thus, it is expected that the performance degradation is not dominant at the SLC-IC scheme. The following section evaluates the proposed schemes by measuring the packet error rate (PER).

\section{Simulation Results} \label{sec:result}

This section presents the results of computer simulation, using MATLAB to assess the gain of the proposed BLC and SLC-IC schemes over a non-combining scheme. The number of transmit and receive antennas, and the number of spatial streams are set to be $N_t = N_r = N_s = 2$, respectively. The maximum number of retransmissions for HARQ is $N = 4$. The channel follows a Rayleigh fading distribution. In each figure, $4$ and $16$ QAM are used to modulate symbols of both desired and interfering signals. A single packet has $400$ uncoded bits. Using a turbo code with the rate $R_c = 0.33, 0.50$ or $0.83$, the coded packet with $400/R_c$ bits is transmitted over $10$ subcarriers with $2$ codewords. A random interleaver is used to permute the bit sequence from the turbo code with a generator polynomials $(7,5)$. This section assumes that channels are perfectly estimated at the receiver.

\begin{figure}[htb]
\centering
\includegraphics[width=3in]{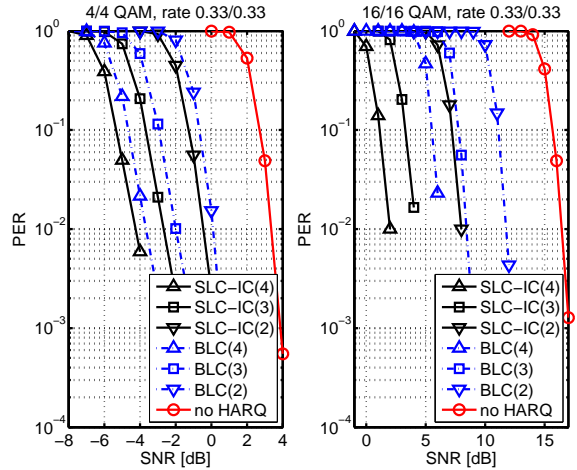}
\caption{The PERs with code rates $0.33$ at SIR $0$ dB are plotted in terms of the number of transmission.}
\label{fig:retransmission}
\end{figure}

\figref{fig:retransmission} shows the PERs of the SLC-IC and BLC schemes with the non-combining scheme in terms of the number of transmissions. The modulation order of both desired and interference signals is $4$ or $16$, and their code rates are $0.33$, respectively. For $4$ QAM, it is observed that the gain of the BLC scheme over the non-combining scheme increases from $4$ to $6$ and $8$ dB, as the number of transmissions increases from $2$ to $3$ and $4$. On the other hand, the SLC-IC scheme achieves the SNR gain of $1$ dB more than the BLC scheme at the same modulation order. When $16$ QAM is used, the gain of the SLC-IC scheme over the BLC scheme is measured up to $4 \sim 5$ dB.

There are three reasons why the SLC-IC scheme outperforms the BLC scheme. First, the BLC scheme is derived under the assumption that the received signal at the $i$th transmission is approximately independent of the received signal at the $(i-k)$th transmission where $k > 0$ in \eqref{eq:BLC}. Second, the sum of the extrinsic LLRs of the desired signal at the BLC scheme propagates the errors caused by the max-log approximation in $(a)$ of \eqref{eq:IASD} by storing the updated LLRs. Lastly, the BLC scheme does not explicitly exploit the decoding results of interference signals at the IASD receiver, although it is fed back to the detector to implicitly improve the extrinsic LLRs of the desired signal. On the other hand, the SLC-IC scheme explicitly exploits the soft information of interference signals to estimate their means and covariances in \eqref{eq:bar_x_I} and \eqref{eq:bar2_x_I}. As the modulation order increases, the approximation error also increase because of the number of constellation points in $(a)$ and its propagation in $(b)$.

\begin{figure}[htb]
\centering
\includegraphics[width=3in]{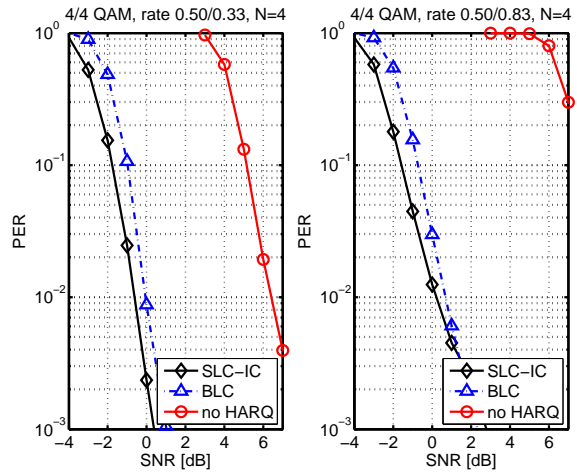}
\caption{The PERs with the code rate $0.50$ for desired signals, and $0.30$ and $0.83$ for interference signals, respectively, are plotted at SIR $0$ dB and $N = 4$.}
\label{fig:cr0.50}
\end{figure}


\figref{fig:cr0.50} plots the PERs of the proposed combining schemes with the code rate of the desired signal $0.50$ where the code rate of the interference signal is changed from $0.33$ to $0.83$. Compared with \figref{fig:retransmission}, it is shown that the PERs are shifted to the right because of the high code rate, $0.50$, of the desired signal. Additionally, as the code rate of the interference signal increases, the non-combining scheme loses a gain by more than $2$ dB. However, the proposed combining schemes are only shifted by less than $1$ dB. This implies that the combining gain obtained from the desired signal partially compensates for the weak decoding gain of the interference signal.



\begin{figure}[htb]
\centering
\includegraphics[width=3in]{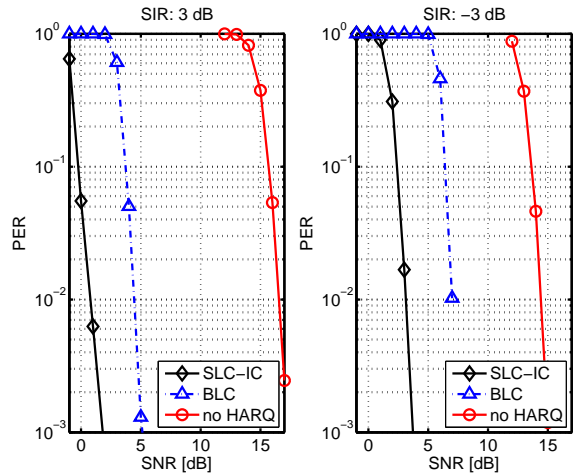}
\caption{The PERs of $N=4$ with $16$ QAM and code rates $0.33$ are plotted at SIR $\pm 3$ dB.}
\label{fig:SIR}
\end{figure}

\figref{fig:SIR} changes the SIR level from $-3$ to $3$ dB under $16$ QAM, the code rate $0.33$ and $N = 4$. When no HARQ is applied, the PER of SIR $-3$ dB outperforms the PER of SIR $3$ dB. This means that the improved \emph{a priori} information of the interference signal is more advantageous than the improved Euclidean distance of the desired signal at the IASD receiver. However, after the proposed HARQ schemes are applied, it is reversed that the PER of SIR $3$ dB becomes better than the PER of SIR $-3$ dB. As the number of transmissions increases up to $4$, it is revealed that the combining gain of the desired signal becomes more important to determine the performance.

\begin{figure}[htb]
\centering
\includegraphics[width=3in]{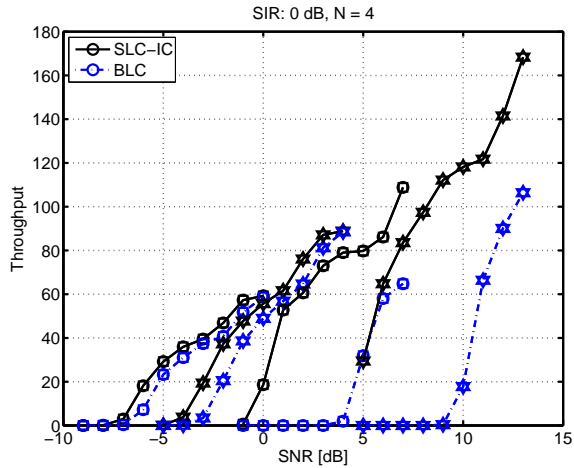}
\caption{The throughput of multiple MCSs is plotted at SIR $0$ dB and $N = 4$. Two modulation order, $4$ and $16$ QAM, and two code rates, $1/3$ and $1/2$ are used.}
\label{fig:MCS}
\end{figure}

\figref{fig:MCS} compares the throughput of both the SLC-IC and BLC schemes in terms of multiple modulation and coding schemes (MCS). In \figref{fig:MCS}, $4$ MCS levels are drawn together. From the leftmost curve, each plot corresponds to $4$ QAM, $0.33$ and $0.50$ code rates, and $16$ QAM, $0.33$ and $0.50$ code rates. \figref{fig:MCS} shows that the SLC-IC scheme still achieves the gain over the BLC scheme with multiple MCS levels not only at the low SNR region but also at the high SNR region.

\section{Conclusion} \label{sec:conclusion}
This paper investigated the problem of combining the desired signals for HARQ at the IASD receiver in a Gaussian interference channel. The desired signal can be requested to retransmit multiple times when HARQ is enabled. However, the interference signal is out of the receiver's control and it varies at each transmission. Since the IASD receiver decodes both the desired and interference signal for its own purpose, the decoding information for the interference signal can be employed for the HARQ operation, too. This paper proposed new schemes to combine only the desired signals in a bit level and in a symbol level, and demonstrated its performance gain over the non-combining scheme. Both the BLC and SLC-IC schemes were shown to offer the improved performance in the presence of interference while requiring some amount of memory units.

\appendices

\section{Interference with Noise Distribution} \label{sec:appendix1}

For notational simplicity, \eqref{eq:residual} can be rewritten by omitting the subscript as
\begin{align}
\mathbf{v} = \mathbf{H}\bracket{\mathbf{x} - \bar{\mathbf{x}}} + \mathbf{n}
\end{align}
where $\mathbf{n} \sim \mathcal{CN}\bracket{0, \mathbf{I}_{N_r}}$. The $n$th element of $\mathbf{x}$ is chosen from the $M$-ary QAM constellation points, each having the symbol probability as
\begin{align}
P(x_n = s) = \prod_{m=1}^{M}\frac{1}{2}\bracket{1 + b_{m,n}^{I}\tanh\bracket{\frac{L_{m,n}^{(A,2,I)}}{2}}}
\end{align}
where $s \in \mathbb{C}$ and $k$ is from $1$ to $2^M$. The bipolar bits $b_{m,n}^{I}$ for all $m = 1, \ldots, M$ correspond to the QAM symbol $s$. This symbol probability is a function of \emph{a posteriori} LLRs of the interference signal at the decoder. Thus, the probability mass function (PMF) of $\mathbf{x}$ varies depending on the decoding performance of the interference signal. When it is not decoded at all, $\mathbf{x}$ is uniformly distributed on the constellation so that $\mathbf{v}$ becomes equal to $\mathbf{n}$. On the other hand, if it is decoded perfectly, $\mathbf{x}$ loses its randomness and becomes constant $\mathbf{s}$. Then, $\mathbf{v} = \mathbf{H}\mathbf{s} + \mathbf{n}$.

In general, $\mathbf{v}$ is the sum of both random variables $\mathbf{x}$ and $\mathbf{n}$. The PDF of $\mathbf{v}$ is calculated by summing the conditional PDF of $\mathbf{n}$ given $\mathbf{x} = \mathbf{s}_k$ as
\begin{align}
f_{\mathbf{v}}\bracket{\mathbf{v}} &= \sum_{\mathbf{s}_k \in \mathbb{C}} f_{\mathbf{v} \mid \mathbf{x}}\bracket{\mathbf{v} - \mathbf{H}\bracket{\mathbf{s}_k - \bar{\mathbf{x}}}} P_{\mathbf{x}}\bracket{\mathbf{x} = \mathbf{s}_k} \\
&\hspace{-0.6cm}= \sum_{\mathbf{s}_k \in \mathbb{C}}  \frac{w_k}{\pi^{N_r}}\exp\bracket{-\norm{\mathbf{v} - \mathbf{H}\bracket{\mathbf{s}_k - \bar{\mathbf{x}}}}^2} \\
&\hspace{-0.6cm}= \sum_{\mathbf{s}_k \in \mathbb{C}}  \frac{1}{\pi^{N_r}}\exp\bracket{-\norm{\mathbf{v} - \mathbf{H}\bracket{\mathbf{s}_k - \bar{\mathbf{x}}}}^2 + \log w_k}
\end{align}
where $w_k = P_{\mathbf{x}}\bracket{\mathbf{x} = \mathbf{s}_k}$. Hence, $\mathbf{v}$ can be modeled with a Gaussian mixture distribution \cite{Kozick00}.

\bibliographystyle{IEEEtran}
\bibliography{IAHARQ}

\end{document}